\documentclass[%
 aip,
 amsmath,amssymb,
 reprint,%
]{revtex4-2}

\usepackage{graphicx}% Include figure files
\usepackage{dcolumn}% Align table columns on decimal point
\usepackage{bm}% bold math
\usepackage{graphicx}% Include figure files
\usepackage{dcolumn}% Align table columns on decimal point
\usepackage{rotating}
\usepackage{lipsum}
\usepackage{xcolor}
\usepackage{bm}% bold math
\usepackage[T1]{fontenc}  
\usepackage{multirow,amsmath,amssymb}
 \usepackage{babel} 
 \usepackage{txfonts}
 \usepackage[percent]{overpic}
 \usepackage{epsfig,epsf,psfrag} 
\usepackage{graphicx} 
\usepackage{pslatex}
\usepackage{epic,eepic} 
\usepackage{color,pstcol}
\usepackage{pstricks} 
\usepackage{fancyhdr} 
\usepackage{tabularx}
\usepackage{physics}
\usepackage{url}
\usepackage{listings}
\usepackage{color} 
\usepackage{caption}
\usepackage{ulem}
\usepackage{balance}
\usepackage{subcaption}
\usepackage{float}
\usepackage[utf8]{inputenc}
\usepackage{comment}
\usepackage{hyperref}
\hypersetup{
 colorlinks=true,
 linkcolor=blue,
 anchorcolor = blue,
 citecolor = blue,
 filecolor = blue,
 urlcolor = blue
}
\usepackage{dsfont}

\graphicspath{{Images/}}
\def \be {\begin{equation}} 
\def \ee {\end{equation}}

\makeatletter
\def\@email#1#2{%
 \endgroup
 \patchcmd{\titleblock@produce}
  {\frontmatter@RRAPformat}
  {\frontmatter@RRAPformat{\produce@RRAP{*#1\href{mailto:#2}{#2}}}\frontmatter@RRAPformat}
  {}{}
}%
\makeatother
\begin{document}

\preprint{AIP/123-QED}

\title[]{Colored $\Delta_T$ noise probes the topological character of edge-modes}
% Force line breaks with \\
\author{Sachiraj Mishra}

 \email{colin.nano@gmail.com}
\author{Colin Benjamin}%
\affiliation{School of Physical Sciences, National Institute of Science Education and Research, HBNI, Jatni-752050, India}
\affiliation{Homi Bhabha National Institute, Training School Complex, AnushaktiNagar, Mumbai, 400094, India }


\begin{abstract}

We investigate colored $\Delta_T$ noise, i.e., finite-frequency $\Delta_T$ noise, as a probe of edge-mode (EM) transport in quantum Hall and quantum spin Hall systems. Colored $\Delta_T$ noise probes finite-frequency nonequilibrium current fluctuations and dynamical transport properties that are often obscured in DC measurements of conductance and noise. Since $\Delta_T$ noise is driven solely by a temperature and voltage bias under zero average charge current conditions, it eliminates current-induced Joule heating and directly probes intrinsic thermal fluctuations. We show that chiral, spin-conserving helical, and spin-flip helical (trivial) EMs exhibit distinct colored $\Delta_T$-noise signatures under appropriate bias protocols. Incorporating energy-dependent scattering through a quantum point contact, we demonstrate that electron-hole asymmetry significantly modifies the finite-frequency spectrum while preserving these distinguishing features. Notably, colored $\Delta_T$ noise exhibits a frequency-dependent sign reversal absent in the corresponding white ($\omega=0$) $\Delta_T$ noise. We further investigate zero-temperature colored quantum shot noise and find that it vanishes identically for chiral EMs, whereas the spin-conserving helical response changes sign with frequency. By contrast, spin-flip helical (trivial) EMs exhibit a positive colored shot-noise spectrum. However, the corresponding colored $\Delta_T$ noise retains its characteristic sign reversal, providing a robust distinction between spin-conserving helical and spin-flip helical (trivial) EM transport. These results establish colored $\Delta_T$ noise as a robust, experimentally accessible, complementary probe for identifying chiral, spin-conserving helical, and spin-flip helical (trivial) EM transport in mesoscopic topological systems.

\end{abstract}

\maketitle

The quantum Hall (QH) effect established topology as a fundamental principle governing electronic transport, with dissipationless conduction through chiral edge states in two-dimensional systems under strong magnetic fields~\cite{PhysRevLett.45.494}. This concept was later extended to the quantum spin Hall (QSH) effect, where spin--orbit coupling generates counterpropagating spin-polarized edge channels~\cite{doi:10.1126/science.1133734, konig2007quantum, roth2009nonlocal}. When spin-flip scattering is absent, these spin-conserving helical EMs are topological and robust against nonmagnetic disorder. However, magnetic-impurity-induced spin-flip scattering can produce spin flip helical (trivial) edge states that mimic topological conductance signatures, making conventional transport measurements insufficient to distinguish between trivial and topological edge transport~\cite{nichele2016edge, PhysRevMaterials.4.104201}. This motivates the search for effective probes beyond conductance.

Quantum noise, arising from current fluctuations at zero and finite frequencies~\cite{noise, PhysRevB.46.12485, balandin2024electronic}, has emerged as a sensitive probe of edge-state transport~\cite{PhysRevB.108.115301}. In particular, $\Delta_T$ noise, generated solely by a temperature gradient in the absence of charge current, isolates intrinsic non-equilibrium fluctuations without Joule heating and therefore can provide a robust diagnostic. Finite-frequency (colored) noise further probes dynamical and energy-resolved aspects of transport inaccessible in DC measurements~\cite{noise, PhysRevB.46.12485, balandin2024electronic}. In quantum systems, the non-commutativity of current operators gives rise to both symmetrized and non-symmetrized finite-frequency noise correlators~\cite{PhysRevB.107.155405}, while the non-symmetrized correlator separately probes emission and absorption processes. Here, we focus on symmetrized colored $\Delta_T$ noise~\cite{lesovik1997detection}. Earlier studies demonstrated that multiterminal conductance measurements can distinguish chiral and spin-conserving helical EM transport \cite{roth2009nonlocal, konig2007quantum}, finite-frequency quantum noise under voltage bias can distinguish chiral, spin-conserving helical, and spin flip helical (trivial) edge transport~\cite{PhysRevB.108.115301}, but such measurements necessarily contain charge-current contributions.

\begin{figure*}
\centering
\includegraphics[scale=0.4]{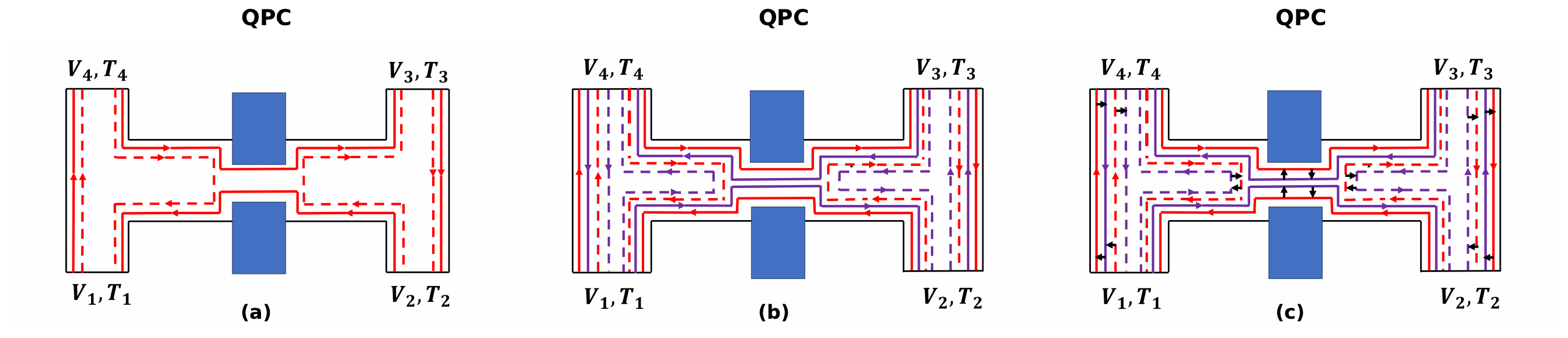}
\caption{(a) Four-terminal quantum Hall device supporting chiral edge transport, (b) four-terminal quantum spin Hall device hosting spin-conserving helical edge states, and (c) four-terminal quantum spin Hall device exhibiting spin-flip helical (trivial) edge states. Yellow solid (dashed) lines depict the transmitted (reflected) EMs. Yellow and purple solid/dashed lines depict spin-up and spin-down EMs. Blue box is the QPC-type constriction with reflection probability $R_0(E)$ and transmission probability $T_0(E)$. Small black arrows in Fig. 1(c) show the spin-flip scattering with probability $P_0$.}
\label{Fig1}
\end{figure*}

$\Delta_T$ noise has been extensively investigated theoretically and experimentally in a variety of mesoscopic systems~\cite{PhysRevLett.127.136801, PhysRevLett.125.086801, PhysRevB.105.195423, PhysRevB.107.075409, PhysRevB.107.245301, PhysRevLett.125.106801, PhysRevB.107.155405, melcer2022absent, popoff2022scattering, lumbroso2018electronic, shein2022electronic, sivre2019electronic, mishra2024andreevreflectionmediateddeltat, PhysRevResearch.7.023321, k95y-7zrb, mishra2025negativespindeltatnoise, shein2024delta, PhysRevB.108.245427}. Experimentally, it has been observed in molecular junctions, mesoscopic circuits, and tunneling devices~\cite{lumbroso2018electronic, shein2024delta, shein2022electronic, sivre2019electronic}, while theoretically it has been connected to thermal-noise bounds, fractional charge, and anyonic statistics~\cite{PhysRevLett.127.136801, PhysRevLett.125.086801, PhysRevB.105.195423, PhysRevB.107.075409, PhysRevB.107.245301, PhysRevB.108.245427}. Other approaches for distinguishing trivial and topological EMs include Hanbury Brown--Twiss correlations~\cite{mani2017probing}, Corbino-geometry noise measurements~\cite{stevens2019noise}, and gate-controlled transport studies~\cite{PhysRevLett.115.036803, PhysRevB.104.045144, PhysRevB.104.195307}. Compared to these approaches, colored $\Delta_T$ noise provides a direct, low-dissipation, and experimentally accessible probe of EM transport.

The remainder of this Letter is organized as follows. We first develop the theoretical framework for quantum noise in the QH configuration, followed by the QSH configuration, including both topological and spin flip helical (trivial) EMs. We then present and discuss the results for colored $\Delta_T$ noise. Finally, we conclude with an analysis of our findings and their possible experimental realization. Further, analysis and details are provided in the Supplemental Material, and the MATHEMATICA code used for computing colored $\Delta_T$ noise is available on GitHub~\cite{github}.

We consider a four-terminal QH configuration with a quantum point contact (QPC) constriction, see Fig. \ref{Fig1}(a). We begin by evaluating the colored quantum noise and subsequently extract the corresponding expression for colored $\Delta_T$ noise. Quantum noise quantifies the fluctuations of current between multiple terminals in a generic mesoscopic configuration. For terminal $\alpha$, the symmetrized current--current correlation function at two different times $t$ and $t'$ is defined as,
\begin{equation}
S_{\alpha\alpha}(t-t') = \frac{1}{2}\Big\langle
\delta \hat{J}_{\alpha}(t)\,\delta \hat{J}_{\beta}(t')
+
\delta \hat{J}_{\beta}(t')\,\delta \hat{J}_{\alpha}(t)
\Big\rangle.
\label{eq:1}
\end{equation}
The fluctuating current operator is given by
$\delta \hat{J}_{\alpha}(t) = \hat{J}_{\alpha}(t) - \langle \hat{J}_{\alpha}(t)\rangle$. Here, $\hat{J}_{\alpha}(t)$ is the total current at terminal $\alpha$ at time $t$ and $\langle \hat{J}_{\alpha}(t) \rangle$ is the average current at terminal $\alpha$ at time $t$.

For a conductor comprising several terminals, the current operator associated with lead $\alpha$ can be written within the scattering formalism as~\cite{noise, PhysRevB.46.12485}
\begin{equation} \label{eq:2}
\begin{split}
\hat{J}_{\alpha}(t) &= \frac{2e}{h} \sum_{\beta,\gamma}\int dE dE'  e^{i(E-E')t/\hbar}\\&
\times \hat{a}^{\dagger}_{\beta}(E) \mathcal{B}_{\beta \gamma}(\alpha;E,E') \hat{a}_{\gamma}(E'),
\end{split}
\end{equation}
where $\beta,\gamma$ label the reservoirs, and the factor of 2 in the prefactor accounts for spin degeneracy. The operators $\hat{a}^{\dagger}_{\beta}(E)$ and $\hat{a}_{\gamma}(E')$ respectively create and annihilate an incoming fermionic state of energies $E$ and $E'$ in leads $\beta$ and $\gamma$. The corresponding occupation number operator is defined as
$
\hat{n}_{\beta}(E)=\hat{a}^{\dagger}_{\beta}(E)\hat{a}_{\gamma}(E')\,\delta_{\beta\gamma}\,\delta(E-E').$ $\mathcal{B}_{\beta \gamma}(\alpha;E,E')$ the kernel, encoding the scattering processes is,
$
\mathcal{B}_{\beta \gamma}(\alpha;E,E') = \delta_{\alpha \beta}\delta_{\alpha \gamma}- \mathcal{M}^{\dagger}_{\alpha \beta}(E) \mathcal{M}_{\alpha \gamma}(E'),$ where $\mathcal{M}_{\alpha \beta}(E)$ is the scattering amplitude from terminal $\beta$ to $\alpha$ at energy $E$, see $s$-matrix for QH in SM in Sec. II of SM.

The average current in terminal $\alpha$ is,
\begin{equation} \label{eq:5}
\begin{split}
\langle J_{\alpha} \rangle =& \frac{2e}{h}\sum_{\beta} \int dE
\Big[\delta_{\alpha \beta}-\mathrm{Tr}(\mathcal{M}_{\alpha \beta}^{\dagger}\mathcal{M}_{\alpha \beta})\Big]f_{\beta}(E - eU_{\beta}),
\end{split}
\end{equation}
where $f_{\beta}(E - eU_{\beta})=[1+e^{(E-eU_{\beta})/k_B T_{\beta}}]^{-1}$ is the Fermi distribution. $U_{\beta}$ and $T_{\beta}$ denote the applied voltage and temperature, respectively at $\beta$. Using ~(\ref{eq:2})–(\ref{eq:5}), current fluctuations and hence the quantum noise correlations can be evaluated \cite{noise, PhysRevB.46.12485}.

 Upon Fourier transforming Eq. (\ref{eq:1}), colored quantum noise is \cite{noise, PhysRevB.46.12485},
\begin{equation}
\begin{split}
2\pi \delta(\omega+\omega') S_{\alpha\alpha}(\omega)
= \frac{1}{2}\Big\langle
\delta \hat{J}_{\alpha}(\omega)\delta \hat{J}_{\beta}(\omega')
+
\delta \hat{J}_{\beta}(\omega')\delta \hat{J}_{\alpha}(\omega)
\Big\rangle.
\end{split}
\label{eq4}
\end{equation}

Following the standard scattering approach, the colored or finite-frequency noise correlations are \cite{noise, PhysRevB.46.12485},

\begin{align}
\begin{split}
&S_{\alpha\alpha}(\omega) = G_0
\sum_{\gamma,\delta}
\int dE \;
\mathcal{B}_{\gamma \delta}(\alpha;E,E_{\omega})\mathcal{B}_{\delta \gamma}(\alpha;E_{\omega},E) 
\\&\Big[
f_{\gamma}(E- eU_{\gamma})\big(1-f_{\delta}(E_{\omega} - eU_{\delta})\big) +
\big(1-f_{\gamma}(E - eU_{\gamma})\big)\\&
\times f_{\delta}(E_{\omega} - eU_{\delta})
\Big],\text{where}\, \,
G_0 = \frac{2e^2}{h},\,\,E_{\omega} = E+\hbar \omega\\
&\text{and}\,\,\mathcal{B}_{\gamma \delta}(\alpha;E,E_{\omega})
= \delta_{\alpha\delta}\delta_{\alpha\gamma}
- \mathcal{M}^{\dagger}_{\alpha\gamma}(E)\,
\mathcal{M}_{\alpha\delta}(E_{\omega}).
\label{eq:12}
\end{split}
\end{align}

 As shown in Sec.~I of the SM~\cite{supp}, the finite-frequency autocorrelation satisfies the symmetry relation
\(S_{\alpha\alpha}(\omega)=S_{\alpha\alpha}(-\omega)\).
Consequently, the symmetrized finite-frequency noise,
\begin{equation}
S_{\alpha\alpha}^{\mathrm{sym}}(\omega)
=
\frac{S_{\alpha\alpha}(\omega)+S_{\alpha\alpha}(-\omega)}{2},
\end{equation}
reduces identically to
\(S_{\alpha\alpha}(\omega)\).
Therefore, for the autocorrelation functions investigated in this work, Eq.~(\ref{eq:12}) directly represents the symmetrized finite-frequency noise.
 
Since we consider an energy-dependent scatterer, the scattering amplitudes explicitly depend on energy. Consequently, the scattering kernel $\mathcal{B}_{\gamma\delta}$ acquires an additional frequency dependence through the factors $\mathcal{M}_{\alpha\delta}(E_{\omega})$ and $\mathcal{M}_{\alpha\gamma}^{\dagger}(E)$. Therefore, unlike the energy-independent case, the finite-frequency noise depends not only on the Fermi-Dirac distribution functions but also on the energy dependence of the scattering probabilities, thereby probing both the occupation of electronic states and the energy-dependent nature of transport.

We evaluate the finite-frequency quantum noise under two different conditions. First, at zero temperature under an applied voltage bias, the quantum noise $S_{\alpha \beta}^U(\omega)$ is decomposed into the finite-frequency vacuum noise, $S_{\alpha\alpha}^{U;\mathrm{va}}(\omega)$, and the zero-temperature colored quantum shot noise, $S_{\alpha\alpha}^{U;\mathrm{sh}}(\omega)$. Second, at finite temperature under the simultaneous application of a voltage bias and a temperature bias, the total finite-frequency noise can be written as
$S_{\alpha\alpha}^{\tau}(\omega)=S_{\alpha\alpha}^{\tau;\mathrm{eq}}(\omega)+S_{\alpha\alpha}^{\tau;\mathrm{va}}(\omega)+S_{\alpha\alpha}^{\tau;\mathrm{b}}(\omega)$,
where $S_{\alpha\alpha}^{\tau;\mathrm{eq}}(\omega)$, $S_{\alpha\alpha}^{\tau;\mathrm{va}}(\omega)$, and $S_{\alpha\alpha}^{\tau;\mathrm{b}}(\omega)$ denote the equilibrium, vacuum, and bias-induced nonequilibrium current-current flutuations in terminal $\alpha$, respectively.

The colored $\Delta_T$ noise is defined as the nonequilibrium finite-frequency current-current correlation under both finite temperature and voltage bias evaluated at a finite thermovoltage ($U_{\mathrm{th}}^{\alpha}$) obtained by imposing net zero charge current transport $\langle J_{\alpha} \rangle = 0$, i.e., 
$\Delta_T^{\alpha\beta}(\omega)= S_{\alpha\alpha}^{\tau;\mathrm{b}}(\omega)\big|_{U=U_{\mathrm{th}}^{\alpha}}$. In the revised letter, the QPC introduces energy-dependent scattering, thereby breaking electron-hole symmetry leading to a finite thermovoltage at zero charge current transported.

The colored $\Delta_T$ noise can be determined experimentally through the following two-step procedure. First, the total finite-frequency current noise $S_{\alpha\alpha}^{\tau}(\omega)$ is measured at zero average charge current, $\langle J_{\alpha}\rangle=0$, in the presence of a finite temperature bias $\Delta T$ and finite thermovoltage $U_{\mathrm{th}}^{\alpha}$. Second, the sum of vacuum-noise contribution, $S_{\alpha\alpha}^{\tau;\mathrm{va}}(\omega)$ and the equilibrium finite-frequency noise, $S_{\alpha\alpha}^{\tau;\mathrm{eq}}(\omega)$, is measured under zero charge current condition, $\langle J_{\alpha}\rangle=0$, and in the absence of any temperature bias ($\Delta T=0$), at the equilibrium temperature. The colored $\Delta_T$ noise is then obtained by subtracting sum of the vacuum-noise and equilibrium-noise contributions from the total measured finite-frequency noise, thereby isolating the nonequilibrium fluctuations generated exclusively by the applied temperature gradient and finite thermovoltage at zero charge current.

To investigate the influence of electron-hole asymmetry and energy-selective transport, we consider energy-dependent scattering. Specifically, we model the constriction as a QPC, whose transmission probability is given by,

\begin{equation}
T_0(E)=\frac{1}{1+\exp\left[-\frac{2\pi(E-E_1)}{\hbar\Omega_x}\right]},
\qquad
E_1=\frac{\hbar\Omega_y}{2},
\end{equation}

where $\hbar\Omega_x$ determines the width of the transmission step and $\hbar\Omega_y$ sets the threshold energy of the QPC. The corresponding reflection probability is $R_0(E)=1-T_0(E)$.

Table~\ref{Table1} summarizes the bias configurations employed. The colored shot noise at terminal `$\alpha$', $S_{\alpha\alpha}^{U}(\omega)$, is evaluated by applying only a voltage bias while maintaining all reservoirs at zero temperature. In contrast, the colored $\Delta_T$ noise is obtained by applying a temperature gradient resulting in a finite corresponding thermovoltage $U_{\mathrm{th}}^{\alpha}$, determined from the zero net charge current condition ($\langle I_2 \rangle = 0$). The reservoir temperatures are parameterized as $\tau_{\pm}=\bar{T}\pm\Delta T/2$, where $\bar{T}$ is the average temperature and $\Delta T$ is the applied temperature difference.

\begin{table*}[t]
\caption{Bias configurations used to calculate the colored shot noise $S_{\alpha \alpha}^{U}(\omega)$ and the colored $\Delta_T$ noise $S_{22}^{\tau}(\omega)$ in Setups 1 and 2. For the temperature-driven case, the applied voltage $U$ is subsequently replaced by the thermovoltage $U_{\mathrm{th}}^{\alpha}$ generated by the temperature gradient under the vanishing average current condition in terminal `$\alpha$'. The temperatures are parameterized as $\tau_{\pm}=\bar{T}\pm\Delta T/2$. $U_{\text{th}}^{\alpha}$ is obtained from the vanishing current in terminal $\alpha$, i.e., $\langle J_{\alpha} \rangle = 0$. }
\label{Table1}
\centering
\begin{tabular}{|c|c|c|c|}
\hline
Configuration & Quantity & Voltages & Temperatures \\
\hline
Setup 1 &
$S_{\alpha \alpha}^{U}(\omega)$ &
$U_1=U_4=U,\;U_2=U_3=0$ &
$\tau_1=\tau_2=\tau_3=\tau_4=0$ \\
\cline{2-4}
&
$S_{\alpha \alpha}^{\tau}(\omega)$ &
$U_1=U_4=U,\;U_2=U_3=0$
($U\rightarrow U_{\mathrm{th}}^{\alpha}$) &
$\tau_1=\tau_4=\tau_{+},\;
\tau_2=\tau_3=\tau_{-}$ \\
\hline
Setup 2 &
$S_{\alpha \alpha}^{U}(\omega)$ &
$U_1=U_3=U,\;U_2=U_4=0$ &
$\tau_1=\tau_2=\tau_3=\tau_4=0$ \\
\cline{2-4}
&
$S_{\alpha \alpha}^{\tau}(\omega)$ &
$U_1=U_3=U,\;U_2=U_4=0$
($U\rightarrow U_{\mathrm{th}}^{\alpha}$) &
$\tau_1=\tau_3=\tau_{+},\;
\tau_2=\tau_4=\tau_{-}$ \\
\hline
\end{tabular}
\end{table*}

For the QH configuration in Fig. \ref{Fig1}(a), the $s$-matrix is derived in Sec. II in SM \cite{supp}, for the QPC, which is an energy-dependent scatterer. To compute the $S_{22}^U(\omega)$, we use Eq.~(\ref{eq:12}), then identify the contributions corresponding to shot noise and vacuum noise.

As derived in Sec. III A 1 of SM, in Setup 1, the expression of $S_{22}^U(\omega)$ given as,

\begin{equation}
\label{eqS22V2}
\begin{split}
S_{22}^{U}(\omega)
&=S_{22}^{U;\mathrm{va}}(\omega)+S_{22}^{U;\mathrm{sh}}(\omega),\\&=2G_0\int_{-\infty}^{\infty}dE
\Big[
\theta(-E)\big(1-\theta(-E_{|\omega|})\big)\\
&
+\big(1-\theta(-E)\big)\theta(-E_{|\omega|})
\Big]=2G_0|\hbar\omega|,
\end{split}
\end{equation}

where, $E_{|\omega|} = E + |\hbar \omega|$.
Here, the shot-noise contribution $S_{22}^{U;\mathrm{sh}}$ vanishes identically as there is no voltage-dependent term in Eq. (\ref{eqS22V2}), and the finite-frequency noise consists entirely of the vacuum noise contribution, which is zero at $\omega=0$ and increases linearly with $|\omega|$.

Similarly, in Setup 1, the thermovoltage $U_{\mathrm{th}}^2$ is evaluated after imposing the average current in terminal 2, i.e., $\langle I_2 \rangle$ to be zero. The relevant Fermi-Dirac distribution functions are given as
$ f_-(E)=\frac{1}{1+e^{E/k_B\tau_-}}$, $f_-(E_{\omega})=\frac{1}{1+e^{(E_{\omega})/k_B\tau_-}}$, $f_+(E)=\frac{1}{1+e^{(E+eU_{\mathrm{th}}^2)/k_B\tau_+}}$, and $f_+(E_{\omega})=\frac{1}{1+e^{(E_{\omega}+eU_{\mathrm{th}}^2)/k_B\tau_+}}$. As derived in Eq. (16) of SM,
the expression of $S_{22}^{\tau}(\omega)$ is,
\begin{equation}\label{eqS22T1}
    \begin{split}
       S_{22}^{\tau}(\omega) &=2G_0|\hbar\omega| + \frac{4G_0\hbar|\omega|}{e^{\hbar|\omega|/(k_B\tau_-)}-1}\\&
       =S_{22}^{\tau;\mathrm{va}} + S_{22}^{\tau;\mathrm{eq}}.
    \end{split}
\end{equation}

 The vacuum noise ($S_{22}^{\tau;\mathrm{va}}$) vanishes in the zero-frequency limit, whereas the equilibrium contribution ($S_{22}^{\tau;\mathrm{eq}}$) disappears as the temperature approaches zero ($\tau_-\rightarrow0$), consistent with the expected limiting behavior of these two noise components. Since the quantum noise contains no term proportional to the applied temperature difference, no colored $\Delta_T$-noise contribution exists in chiral QH regime, implying $\Delta_T^{22}(\omega)=0$. By symmetry, an identical result is obtained for terminal 4, i.e., $\Delta_T^{44}(\omega)=0$.

Finally, as in Setup 1, one can explicitly show for Setup 2 that both the shot noise-like autocorrelation $S_{22}^{\text{sh}}(\omega)$ and the $\Delta_T$ noise autocorrelation $\Delta_T^{22}(\omega)$ vanish at finite frequencies in the case of chiral edge transport.

Next, we discuss colored $\Delta_T$ noise in a four-terminal QSH configuration considering either spin-conserving helical as in Fig. \ref{Fig1}(b) or spin flip helical (trivial) EMs as in Fig. \ref{Fig1}(c). For trivial EMs, the spin-flip scattering can occur with probability $P_0$. $P_0 = 0$ reduces to the helical case. As in the QH case, we first evaluate the finite-frequency quantum noise correlations such as $S_{\alpha \alpha}^U(\omega)$ and $S_{\alpha \alpha}^{\tau}(\omega)$ and then extract the corresponding expressions for colored zero-temperature quantum shot noise ($S_{\alpha \alpha}^{\text{sh}}(\omega)$) and colored $\Delta_T$ noise ($\Delta_T^{\alpha \alpha}(\omega)$). In a QSH Setup, noise correlations are described by resolving the currents into their spin components. The colored quantum noise correlation between currents in terminals $\alpha$ and $\beta$ with spin $\sigma$ and $\sigma'$ is defined as \cite{PhysRevB.75.085328}:
\begin{equation}
\begin{split}
S_{\alpha\alpha}^{\sigma \sigma'}(t-t') &=
\frac{1}{2}\Big\langle
\delta \hat{J}_{\alpha}^{\sigma}(t)\,
\delta \hat{J}_{\beta}^{\sigma'}(t')
+
\delta \hat{J}_{\beta}^{\sigma'}(t')\,
\delta \hat{J}_{\alpha}^{\sigma}(t)
\Big\rangle,\\& \text{with,} \,\,\delta \hat{J}_{\alpha}^{\sigma}(t)
=
\hat{J}_{\alpha}^{\sigma}(t)
-
\langle \hat{J}_{\alpha}^{\sigma}(t)\rangle.
\end{split}
\end{equation}

Here, $\hat{J}_{\alpha}^{\sigma}(t)$ denotes the spin-resolved current operator, while $\langle \hat{J}_{\alpha}^{\sigma}(t) \rangle$ represents the corresponding average current in terminal $\alpha$ for spin $\sigma$ at time $t$. Spin-polarized current $\hat{J}_{\alpha}^{\sigma}(t)$ is given by \cite{PhysRevB.75.085328},
\begin{equation} \label{eq:49}
\begin{split}
\hat{J}_{\alpha}^{\sigma}(t) = \frac{e}{h} \int \int dE \, dE' \, e^{i(E-E')t/\hbar} \\
\times \Big(\hat{a}_{\alpha}^{\sigma \dagger}(E)\hat{a}_{\alpha}^{\sigma}(E') - \hat{b}_{\alpha}^{\sigma \dagger}(E)\hat{b}_{\alpha}^{\sigma}(E')\Big).
\end{split}
\end{equation}

The operators $\hat{a}_{\alpha}^{\sigma\dagger}(E)$ and $\hat{a}_{\alpha}^{\sigma}(E)$ respectively describe the creation and annihilation of an incident electron with spin $\sigma$ and energy $E$ in lead $\alpha$, whereas $\hat{b}_{\alpha}^{\sigma\dagger}(E)$ and $\hat{b}_{\alpha}^{\sigma}(E)$ represent the corresponding outgoing states. The expectation value of the spin-resolved current at terminal $\alpha$ is then obtained as~\cite{PhysRevB.75.085328},
\begin{equation}
\langle J_{\alpha}^{\sigma} \rangle = \frac{e}{h}\sum_{\beta}\int dE \Big(\delta_{\alpha \beta} - \mathrm{Tr}\big(\mathcal{M}_{\alpha \beta}^{\sigma \dagger}(E) \mathcal{M}_{\alpha \beta}^{\sigma}\big)(E)\Big) f_{\beta}(E - eU_{\beta}).
\end{equation}

The spin-dependent scattering matrix is introduced as
$\mathcal{M}_{\alpha \beta}^{\sigma} = \sum_{\rho \in \{\uparrow, \downarrow\}} \mathcal{M}_{\alpha \beta}^{\rho \sigma}$,
where $\mathcal{M}_{\alpha \beta}^{\rho \sigma}$ denotes the scattering amplitude associated with a carrier entering from terminal $\beta$ with spin $\sigma$ and exiting at terminal $\alpha$ with spin $\rho$. The scattering matrix elements $\mathcal{M}_{\alpha \beta}^{\rho \sigma}$ for the spin-conserving helical EM and spin-flip based trivial EMs are given in Eqs. (10) and (11) of SM \cite{supp} in Sec. II. The function $f_{\beta}(E - eU_{\beta})$ is the Fermi distribution defined below Eq.~(\ref{eq:5}).

On Fourier transforming the time-domain correlator similar to Eq. (\ref{eq4}), the generic expression for spin-resolved finite-frequency noise is~\cite{PhysRevB.75.085328},
\begin{flalign}
\begin{split}
&S_{\alpha\alpha}^{\sigma \sigma'}(\omega)
=
\frac{G_0}{2}
\sum_{\mu,\nu \in \{\uparrow,\downarrow\}}
\sum_{\gamma,\delta}
\int dE \;
\mathrm{Tr}\Big[
\mathcal{B}_{\gamma \delta}^{\mu \nu}
(\alpha, \sigma;E,E_{\omega}) \\
&\qquad \times
\mathcal{B}_{\delta \gamma}^{\nu \mu}
(\beta, \sigma';E_{\omega},E)
\Big] 
\Big[
f_{\gamma}(E - eU_{\gamma})
\big(1 - f_{\delta}(E_{\omega} - eU_{\delta})\big)
\\&+
\big(1 - f_{\gamma}(E - eU_{\gamma})\big)f_{\delta}(E_{\omega} - eU_{\delta})
\Big],\,\,\text{where}\\
&\mathcal{B}_{\gamma \delta}^{\mu\nu}(\alpha, \sigma;E,E_{\omega})
= \delta_{\alpha\delta}\delta_{\alpha\gamma}\delta_{\sigma\mu}\delta_{\sigma\nu}
- 
\mathcal{M}^{\sigma\mu;\dagger}_{\alpha\gamma}(E)\,
\mathcal{M}_{\alpha\delta}^{\sigma\nu}(E_{\omega}).
\label{eq:13}
\end{split}
\end{flalign}

\begin{figure*}
     \centering
     \begin{subfigure}[b]{0.45\textwidth}
         \centering
         \includegraphics[width=\textwidth]{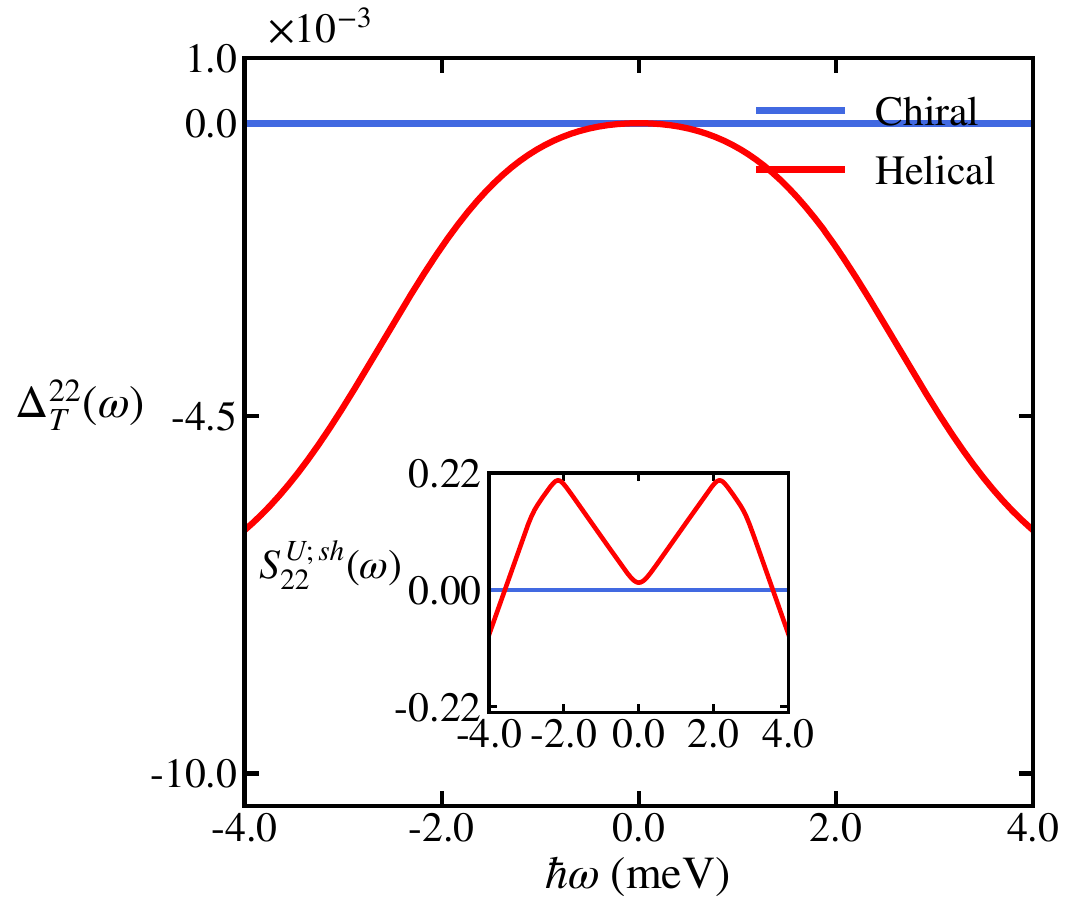}
         \label{fig:3(a)}
         \caption{}
     \end{subfigure}
     \hspace{0.07cm}
     \begin{subfigure}[b]{0.45\textwidth}
         \centering
         \includegraphics[width=\textwidth]{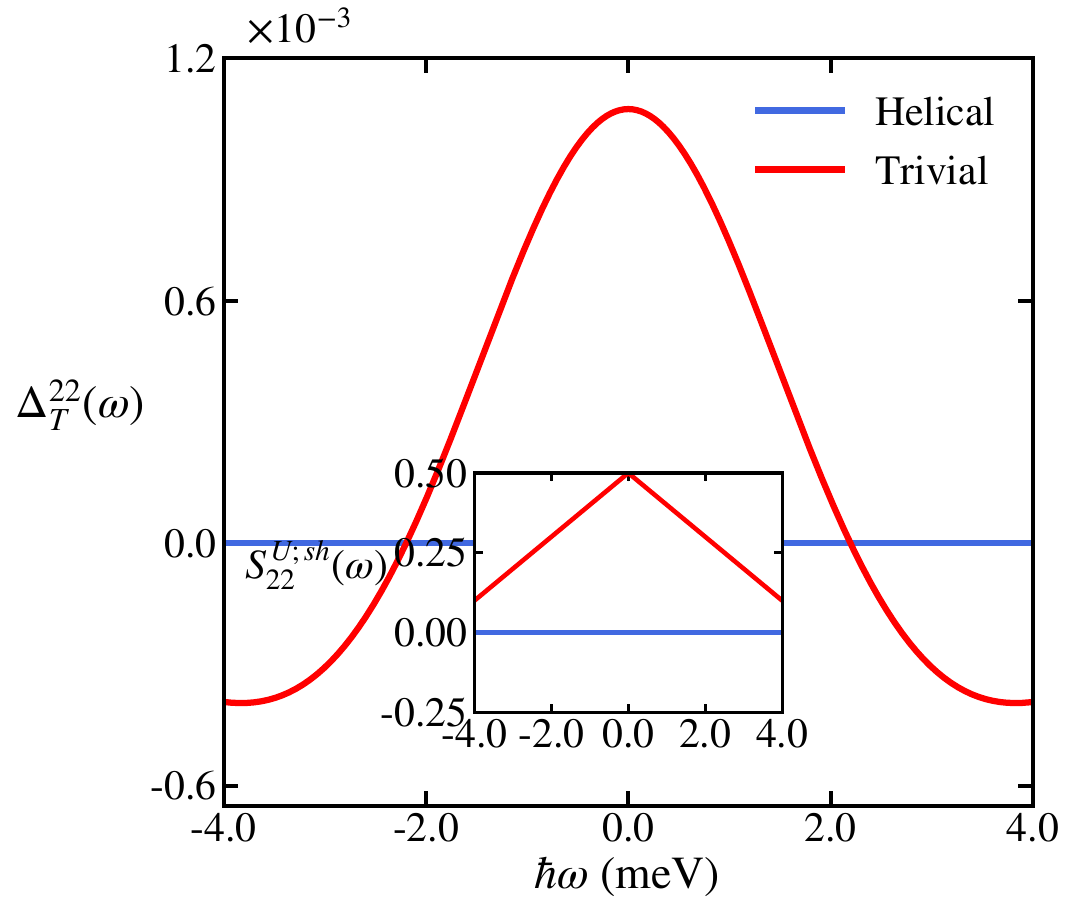}
         \label{fig:3(d)}
         \caption{}
     \end{subfigure}
        \caption{ (a) Colored $\Delta_T$ noise autocorrelation $\Delta_T^{22}(\omega)$ in Setup 1 i.e., $\tau_{1(4)} = \tau_+, \tau_{2(3)} = \tau_-$ at $U_1 = U_4 = U_{\mathrm{th}}^2$ for $\langle I_2 \rangle = 0$, and $U_2 = U_3 = 0$ in units of $G_0 k_B \bar{T}$. Parameters: $\Delta T = 0.1 \bar{T}$ for $\Delta_T$ noise with $\bar{T} = 10 K$ (Inset: Zero-temperature colored quantum shot noise autocorrelation $S_{22}^{U;\mathrm{sh}}(\omega)$ in Setup 1, i.e., $U_1 = U_4 = U, U_2 = U_3 = 0$ at $\tau_1 = \tau_2 = \tau_3 = \tau_4 = 0$ in units of $G_0 eU$), (b) colored $\Delta_T$ noise autocorrelation $\Delta_T^{22}(\omega)$ in Setup 2, i.e., $\tau_{1(3)} = \tau_+, \tau_{2(4)} = \tau_-$ at $U_1 = U_3 = U_{\mathrm{th}}^2$ for $\langle I_2 \rangle = 0$, and $U_2 = U_4 = 0$ in units of $G_0 k_B \bar{T}$ (Inset: Zero-temperature colored quantum shot noise autocorrelation $S_{22}^{U;\mathrm{sh}}(\omega)$ in Setup 2 i.e., $U_1 = U_3 = U, U_2 = U_4 = 0$ at $\tau_1 = \tau_2 = \tau_3 = \tau_4 = 0$  in units of $G_0 eU$.) Parameters: $eU = 5meV$, $\hbar \Omega_y = 4.3meV, \hbar \Omega_x = 0.1 \hbar \Omega_y$. }
         \label{fig:10}
       \end{figure*}

Quantum noise is obtained by summing over all spin components: $S_{\alpha \alpha}(\omega) = \sum_{\sigma, \sigma' \in \{\uparrow, \downarrow \}}  S_{\alpha \alpha}^{\sigma \sigma'}$. Just like QH case, here also for the autocorrelations $S_{\alpha \alpha}(\omega) = S_{\alpha \alpha}(-\omega)$ is satisifed. We now discuss the colored quantum noise at zero temperature for spin-conserving helical edge states, focusing on Setup 1. Using Eq.~(\ref{eq:13}), the colored quantum noise $S_{22}^U(\omega)$ at zero-temperature can be decomposed into spin-resolved contributions as $S_{22}^U(\omega) = \sum_{\sigma,\sigma' \in {\uparrow,\downarrow}} S_{22}^{\sigma \sigma';U}(\omega)$, where $S_{22}^{\sigma \sigma';U}(\omega)$ represent the individual spin-resolved, frequency-dependent components. Similarly, autocorrelations in other terminals, such as $S_{44}^U(\omega)$. Substituting Eq.~(\ref{eq:13}) along with the scattering matrix corresponding to the QSH configuration with spin-conserving helical EMs (see, Eq.~(11) of Sec.~II B of SM \cite{supp}), and following the derivation presented in Sec.~III B of SM \cite{supp}, we find that $S_{22}^U(\omega)$ consists of two distinct contributions: a shot noise-like term and a frequency-dependent vacuum noise term. The mixed-spin correlations, $S_{22}^{\uparrow\downarrow;U}(\omega)$ and $S_{22}^{\downarrow\uparrow;U}(\omega)$, are zero in the absence of spin-flip processes, see Eq. (20) of Sec. III B of SM \cite{supp}. Specifically, the shot noise contribution $S_{22}^{\text{sh}}(\omega)$ originates from partitioning at the scatterer and is given as

\begin{equation}
\begin{split}
S_{22}^{U;\mathrm{sh}}(\omega)
=&\,
\frac{G_0}{2}\Bigg[
\theta(eU-|\hbar\omega|)
\left(
\int_{0}^{eU-|\hbar\omega|}\!\!\mathcal{X}(E)\,dE
+
\int_{-|\hbar\omega|}^{0}\!\!\mathcal{Y}(E)\,dE
\right)
\\
&
+
\theta(|\hbar\omega|-eU)
\int_{-|\hbar\omega|}^{eU-|\hbar\omega|}
\mathcal{Y}(E)\,dE
\Bigg].
\end{split}
\label{eq:41c}
\end{equation}

where, $\mathcal{X}(E) = T_0(E)R_0(E_{|\omega|}) + T_0(E_{|\omega|})R_0(E)$ and $\mathcal{Y}(E) = T_0(E)R_0(E_{|\omega|}) - T_0(E_{|\omega|})R_0(E)$. The same qualitative behavior is obtained for the autocorrelation measured at terminal~4. Consequently, the existence of a finite colored shot-noise autocorrelation, $S_{22}^{U;\mathrm{sh}}(\omega)$, in the helical EM configuration provides a clear distinction from chiral EM case. In particular, the colored shot-noise autocorrelation identically vanishes for chiral EMs, i.e., $S_{22}^{U;\mathrm{sh}}(\omega)=0$ [see, Eq.~(\ref{eqS22V2})], whereas it remains finite for spin-conserving helical EMs and exhibits a pronounced frequency dependence. For the results presented in Fig.~\ref{fig:10}(a), we consider the regime $eU>|\hbar\omega|$ with a fixed bias of $eU=5\,\mathrm{meV}$. As shown in the inset of Fig.~\ref{fig:10}(a), $S_{22}^{U;\mathrm{sh}}(\omega)$ has a small positive value around zero frequency. As $|\hbar\omega|$ increases, the shot noise initially increases and reaches a pronounced maximum at finite frequency. Upon further increasing the frequency, its magnitude decreases continuously and eventually undergoes a sign reversal. This behavior can be understood directly from Eq.~(\ref{eq:41c}). In the regime $eU>|\hbar\omega|$, only the first two terms contribute, while the third term identically vanishes. The first term is strictly positive because both the scattering factor,
$
\mathcal{X}(E),
$
and the corresponding integration interval are always positive. Consequently, it always provides a positive contribution to the shot noise. The second term $\mathcal{Y}(E)
$, which contains the difference of scattering probabilities,
can change sign as the frequency increases, see Sec. V of SM \cite{supp}. The competition between the positive contribution from the first term and the sign-changing contribution from the second term is therefore responsible for the observed non-monotonic behavior and the eventual sign reversal of the colored shot-noise spectrum. This sign reversal is absent for chiral EMs, where the colored shot-noise autocorrelation vanishes identically, thereby providing a robust and experimentally accessible signature for distinguishing helical edge transport from chiral edge transport.

We next evaluate the colored $\Delta_T$ noise in a QSH (Fig. \ref{Fig1}(b)) sample for Setup 1, with $\tau_{1(4)} = \tau_+, \tau_{2(3)} = \tau_-$. As in voltage-bias case, the the quantum noise correlation can be decomposed into spin-resolved components as $S_{22}^{\tau}(\omega) = \sum_{\sigma, \sigma' \in \{\uparrow, \downarrow \}}S_{22}^{\sigma \sigma'; \tau}(\omega)$. Using Eq. (\ref{eq:13}), one can derive the expressions for $S_{22}^{\sigma \sigma'; \tau}(\omega)$ and $S_{22}^{\tau}(\omega)$ can be decomposed as sum of $S_{22}^{\tau; \mathrm{eq}}(\omega)$ and $  S_{22}^{\tau;\mathrm{va}}(\omega)$ and $ \Delta_T^{22}(\omega)$ (see, Sec. II B 2 of SM \cite{supp}). In the linear-response regime, $\langle I_2 \rangle = 0$ yields the thermovoltage $U_{\mathrm{th}}^{2}$, whose derivation is presented in Sec.~IV A of the SM \cite{supp}. The resulting expression is 

\begin{equation}
U_{\mathrm{th}}^2
=\left(
\frac{\int_{-\infty}^{\infty} dE(1-2T_0(E))E\left(-\frac{\partial f_0}{\partial E}\right)}
{\int_{-\infty}^{\infty} dE T_0(E)\left(-\frac{\partial f_0}{\partial E}\right)}
\right)
\frac{\Delta T}{2e\bar{T}}.
\end{equation}
$\Delta_T^{22}$ is subsequently evaluated by setting the applied voltage equal to the corresponding thermovoltage, i.e., $U_{\text{th}}^2$. 

An analogous procedure is followed for terminal 4. The thermovoltage $U_{\mathrm{th}}^{4}$ is obtained by enforcing the condition $\langle I_4\rangle=0$. The derivation is provided in Sec.~II of the Supplemental Material. The thermovoltage also serves as a clear indicator of the underlying edge-state configuration. Specifically, it acquires a finite value for spin-conserving helical EMs, whereas it remains exactly zero for chiral EMs, as illustrated in Fig.~1 of the SM.
 As derived in Sec. III B 2 of SM, the expression for $\Delta_T$ noise is 

 \begin{equation}
\begin{split}
\Delta_T^{22}(\omega)
=&\,
\frac{G_0}{2}
\int_{-\infty}^{\infty} dE
\Bigg\{
\mathcal{X}(E)
\Bigg[
(eU_{\mathrm{th}}^2)^2
\frac{\partial f_0(E)}{\partial E}
\frac{\partial f_0(E_{\omega})}{\partial E}
\\
&
\qquad
-
eU_{\mathrm{th}}^2\Delta T
\left(
\frac{\partial f_0(E)}{\partial E}
\frac{\partial f_0(E_{\omega})}{\partial \bar{T}}
+
\frac{\partial f_0(E)}{\partial \bar{T}}
\frac{\partial f_0(E_{\omega})}{\partial E}
\right)
\\
&
\qquad
+
(\Delta T)^2
\frac{\partial f_0(E)}{\partial \bar{T}}
\frac{\partial f_0(E_{\omega})}{\partial \bar{T}}
\Bigg]
\\
&
+
\mathcal{Y}(E)
\Bigg[
-eU_{\mathrm{th}}^2
\left(
f_0(E)
\frac{\partial f_0(E_{\omega})}{\partial E}
-
f_0(E_{\omega})
\frac{\partial f_0(E)}{\partial E}
\right)
\\
&
\qquad
+
\Delta T
\left(
f_0(E)
\frac{\partial f_0(E_{\omega})}{\partial \bar{T}}
-
f_0(E_{\omega})
\frac{\partial f_0(E)}{\partial \bar{T}}
\right)
\Bigg]
\Bigg\}.
\end{split}
\label{eq:41b}
\end{equation}

where $f_0 = \frac{1}{1+e^{\frac{E}{k_B \bar{T}}}}$. In the chiral case, $\Delta_T^{22}(\omega)$ vanishes, while for the spin-conserving helical case it remains finite. As shown in Fig.~\ref{fig:10}(a), $\Delta_T^{22}(\omega)$ exhibits a non-monotonic dependence on frequency and is symmetric with respect to $\omega$, as expected. At zero frequency, $\Delta_T^{22}(\omega)$ attains its peak value. With increasing frequency, it decreases and changes sign to become negative. Here, the $\Delta_T$ noise remains almost always negative. This observation distinguishes chiral and helical edge states through temperature-biased noise measurements.

Colored shot noise and colored $\Delta_T$ noise for spin flip helical (trivial) edge states are compared only in Setup 2. We don't do so in Setup 1, as there is only quantitative difference between spin-conserving helical and spin-flip helical (trivial) edge states. In Setup 2, for spin flip helical (trivial) EMs, the zero-temperature quantum noise autocorrelation $S_{22}^{U}(\omega)$ is derived in Eq. (65) in Sec. III C of SM \cite{supp}. In Eq. (65), the coefficient proportional to $\hbar \omega$ is the vacuum noise due to finite frequency, i.e., $S_{22}^{U; \mathrm{va}} = 2 G_0(1-P_0) \hbar \omega$, and the term proportional to $(eU-|\hbar \omega|)$ is shot noise, i.e., $S_{22}^{U;\text{sh}}(\omega) = 2G_0 P_0(1-P_0) (eU-|\hbar \omega|) \theta(eU-|\hbar \omega|).$ We see that $S_{22}^{U;\text{sh}} (\omega)$ vanishes at $ P_0 = 0$ (i.e., for spin-conserving helical), but is finite for $ P_0 \neq 0$ (spin flip based helical (trivial)). We also see that $S_{22}^{U;\text{sh}}$ reduces linearly with $|\hbar \omega|$ for $eU>|\hbar \omega|$ (see, inset of Fig. \ref{fig:10}(b) above), and it being finite helps in distinguishing helical from spin-flip based helical (trivial) EM qualitatively.

The expression for $\Delta_T^{22}(\omega)$ for spin-flip based helical (trivial) EMs is (see, SM Sec. III C for full derivation), given below,
\begin{equation}\label{eq:47}
\begin{split}
\Delta_T^{22}(\omega)
=
2G_0 P_0(1-P_0)(\Delta T)^2
\int_{-\infty}^{\infty} dE
\Bigg[
&
\frac{\partial f_0(E)}{\partial \bar{T}}
\frac{\partial f_0(E_{\omega})}{\partial \bar{T}}
\Bigg].
\end{split}
\end{equation}

The autocorrelation $\Delta_T^{22}(\omega)$ vanishes for spin-conserving helical EMs ($ P_0 = 0$), but is finite for spin-flip helical (trivial) EMs ($ P_0 \neq 0$), as depicted in Fig. \ref{fig:10}(b) above. Similar to Fig.~\ref{fig:10}(a), $\Delta_T$ noise autocorrelation $\Delta_T^{22}(\omega)$ also remains symmetric with respect to $\omega$, as in Fig.~\ref{fig:10}(b). In white noise limit ($\omega \to 0$), $\Delta_T^{22}(\omega)$ reaches its maximum. As frequency increases, it decreases, turns negative, develops a pronounced dip, and then rises again, eventually approaching zero at higher frequencies. Thus, colored $\Delta_T$ noise differentiates spin-conserving helical and spin flip based helical (trivial) EMs. We have summarized the results for zero-temperature colored quantum shot noise and colored $\Delta_T$ noise across both the setups in Table \ref{Table2}.

\begin{table}[H]
\caption{Magnitude and sign of colored shot noise and colored $\Delta_T$ noise in Setups 1 and 2.}
\label{Table2}
\centering
\resizebox{8.90cm}{!}{%
\begin{tabular}{c|cccccccccccc|cccccccccccc|}
\cline{2-25}
 & \multicolumn{12}{c|}{Setup 1} & \multicolumn{12}{c|}{Setup 2} \\ \cline{2-25} 
 & \multicolumn{5}{c|}{$S_{22}^{U;\text{sh}}(\omega)$} & \multicolumn{7}{c|}{$\Delta_T^{22}(\omega)$} & \multicolumn{5}{c|}{$S_{22}^{U;\text{sh}}(\omega)$} & \multicolumn{7}{c|}{$\Delta_T^{22}(\omega)$} \\ \hline
\multicolumn{1}{|c|}{Chiral}  & \multicolumn{5}{c|}{Zero}     & \multicolumn{7}{c|}{Zero}                  & \multicolumn{5}{c|}{Zero}       & \multicolumn{7}{c|}{Zero}                    \\ \hline
\multicolumn{1}{|c|}{Spin-conserving Helical} & \multicolumn{5}{c|}{Changes Sign} & \multicolumn{7}{c|}{\begin{tabular}{c}
Changes sign
\end{tabular}} & \multicolumn{5}{c|}{Zero}     & \multicolumn{7}{c|}{Zero}                  \\ \hline
\multicolumn{1}{|c|}{Spin-flip Helical (Trivial)} & \multicolumn{5}{c|}{Changes Sign}       & \multicolumn{7}{c|}{\begin{tabular}{c}
Changes sign
\end{tabular}}                    & \multicolumn{5}{c|}{Positive or Zero} & \multicolumn{7}{c|}{\begin{tabular}{c}
Changes sign
\end{tabular}} \\ \hline
\end{tabular}%
}
\end{table}

In this Letter, we have presented an experimentally accessible framework that uses colored $\Delta_T$ noise to differentiate between topological (chiral and spin-conserving helical) and spin flip helical (trivial) EMs.

We consider frequencies in the range
$\frac{-4 meV}{\hbar} \leq \omega \leq \frac{4 meV}{\hbar}$, as shown in Fig.~\ref{fig:10}. The characteristic sign reversal in the colored shot noise or colored $\Delta_T$ noise spectra occurs over the finite frequency range corresponding to approximately $\hbar \omega$ of the order of $\mathrm{meV}$, which translates to frequencies of the order of 1-10$~\mathrm{GHz}$. Although experimental measurements of colored $\Delta_T$ noise in molecular junctions have so far been limited to lower frequencies~\cite{shein2024delta}, finite-frequency quantum shot noise has already been measured in mesoscopic systems in the GHz--THz regime~\cite{PhysRevLett.99.236803, bisognin2019microwave, jompol2015detecting, PhysRevLett.116.227401}. Therefore, the proposed setups can be adapted to probe colored $\Delta_T$ noise experimentally at high frequencies.

Experimentally, colored shot noise, typically in the GHz \cite{PhysRevLett.99.236803, bisognin2019microwave} or THz \cite{jompol2015detecting, PhysRevLett.116.227401} regime, in mesoscopic conductors is detected using microwave measurement techniques~\cite{PhysRevLett.99.236803, bisognin2019microwave, jompol2015detecting, PhysRevLett.116.227401}. In such experiments, the mesoscopic device is connected to impedance-matched transmission lines that are designed to carry high-frequency signals efficiently. These transmission lines collect the noise signals emitted by the device and guide them toward the detection circuitry. Generally the emitted noise is extremely weak, it is first amplified using cryogenic low-noise amplifiers placed at low temperatures to minimize additional background noise. The noise power is then measured using calibrated detectors that convert the high-frequency signal into a measurable output. To extract the intrinsic noise of the device, bias-modulation techniques are employed so that background contributions can be removed reliably.

A central outcome of this work is the qualitative distinction between zero-temperature colored quantum shot noise and colored $\Delta_T$ noise as probes of edge-state transport. For chiral EMs, both quantities vanish identically. In contrast, for spin-conserving helical EMs, the colored quantum shot noise undergoes a frequency-dependent sign reversal, whereas the colored $\Delta_T$ noise remains predominantly negative over the entire frequency range considered. For spin-flip helical (trivial) EMs, the colored quantum shot noise differs from its topological counterpart only in the regime $|\hbar\omega|>eU$, making the distinction dependent on the relative magnitudes of the applied voltage and frequency. The colored $\Delta_T$ noise, however, clearly distinguishes spin-conserving helical and spin-flip helical (trivial) edge transport over the entire frequency range. These results establish colored $\Delta_T$ noise as a robust and experimentally accessible probe for identifying chiral, spin-conserving helical, and spin-flip helical (trivial) EMs. Experimentally, the colored $\Delta_T$ noise is obtained by subtracting the equilibrium finite-frequency noise, measured at $\Delta T=0$, from the total finite-frequency noise measured under an applied temperature bias. 

\textbf{Supplemental Material:} See the Supplemental Material (SM) \cite{supp} for the proof of the symmetry of the quantum-noise autocorrelation; derivations of the scattering matrices for the chiral QH, spin-conserving helical QSH, and spin-flip-induced helical (trivial) QSH configurations; detailed derivations of the zero-temperature colored quantum shot noise and colored $\Delta_T$ noise; the derivation and analysis of the thermovoltage; physical explanations of the results presented in this paper; and a discussion of the zero-temperature colored quantum shot noise and colored $\Delta_T$ noise in the high-frequency regime.

\textbf{Acknowledgment:} The authors thank the two anonymous Referees for their helpful comments and valuable suggestions, which have significantly improved the quality and presentation of the manuscript.

% Finally, we add the Mathematica code to calculate $\mathcal{D}_T$ noise in Appendix \ref{code}.

\bibliography{apssamp}

\end{document}